\documentstyle[prl,aps,twocolumn]{revtex}
\begin{document}
\draft

\twocolumn[\hsize\textwidth\columnwidth\hsize\csname
@twocolumnfalse\endcsname
\begin{flushright}
hep-th/0008037
\end{flushright}
\vspace{1.0cm}
\title{Coarse graining and renormalization}
\draft
\author{Ji-Feng Yang\thanks{email:jfyang@fudan.edu.cn}}
\address{Department of Physics, East China Normal University,
Shanghai 200062, P. R. China}
\date{October 20 2001}
\maketitle

\begin{abstract}
Formulating the QFT's as coarse grained 'low' energy sectors of a
postulated complete quantum theory of everything with the 'high'
energy modes integrated out or 'clustering' into 'low' energy
objects, we can evaluate the Feynman amplitudes by solving a
series of natural differential equations which automatically
dissolves the necessity of infinity subtraction and the associated
subtleties. This new strategy has direct implications to the
scheme dependence problem.
\end{abstract}

 ]
\narrowtext

Conventionally, a quantum field theory is formally established first and
then regularized by hand and finally renormalized in order to arrive at
finite predictions to confront experimental outcomes. This 'trilogy' has
been successfully applied for over half a century. However, it is 
undeniable
that there are some inharmonious and even irrational 'notes' in this
'trilogy' that has also been dissatisfying many physicists for half a
century, e.g., to treat an infinity as an infinitesimal, or to discard 
an
explicitly infinite part while keeping the finite piece, no matter how 
one
argued in favor of doing so, even within the beautiful scenario due to 
Wilson%
\cite{Wilson}. The widely used expression for the beta function of a $%
U\left( 1\right) $ coupling $\beta \left( \alpha \right) \equiv
-\frac \alpha {Z_\alpha }\mu \partial _\mu Z_\alpha $ is derived
from
\begin{eqnarray}
\mu \frac d{%
d\mu }\alpha _B(=Z_\alpha \alpha)=0\Rightarrow \left( \mu
\partial _\mu Z_\alpha \right) \alpha +Z_\alpha \mu
\partial _\mu \alpha =0
\end{eqnarray}
which should be corrected as\cite{RMP}
\begin{eqnarray}
\mu \frac d{d\mu }%
&&\left( Z_\alpha \alpha \right) =0\Rightarrow Z_\alpha \mu
\partial _\mu \alpha +\alpha \mu \partial _\mu Z_\alpha +\alpha
\mu \partial _\mu \alpha
\partial _\alpha Z_\alpha =0\nonumber \\
&&\Rightarrow \beta \left( \alpha \right) =-\frac{%
\alpha \mu \partial _\mu Z_\alpha }{Z_\alpha +\alpha \partial _\alpha
Z_\alpha },
\end{eqnarray}
because it is illogical to take the coupling in $Z_\alpha \left(
\Lambda ,\mu ,\alpha \right) $ as the bare one during
differentiation and to identify it as the renormalized one after
the differentiation. Such a switch of attitude can not be
reasonably accepted. Obviously, all these unsatisfactory
subtleties are due to the inevitable appearance of ultra-violet
divergence in conventional methods.

In this letter, we wish to show a simple way to get rid of these
unsatisfactory subtleties by adopting a natural point of view that the
conventional QFT should be replaced by a complete quantum theory of
everything (QTOE) which contain the correct higher energy structures as
well. The low energy objects (fields or particles) effectively emerged 
as
some kind of 'clusters'\cite{Cluster} or collective modes of the quanta 
that
are extremely small in sizes. Then the low energy physics are defined 
by the
coarse grained low energy sectors of QTOE with the extremely short 
distance
processes integrated out\cite{Coarse}. In this understanding, high 
energy
modes' contributions are physically suppressed by the clustering 
mechanism
defined in QTOE (unknown to us) rather than cut off by hand, a 
refinement of
the Wilsonian scenario. This picture naturally necessitates the 
presence of
a set of parameters to characterize the high energy modes in 
the 'clusters'
(which will be collectively denoted as $\left\{ \sigma \right\} $) and 
the
clustering mechanism (which will be denote as a threshold scale $\bar
{\mu}$
for clustering of underlying high energy modes in an appropriate QFT).
Technically, it is these constants that suppressed the high energy 
modes in
QTOE while keep the 'effective' quanta dominant. For the coarse 
graining or
emergence scenario to be effective, the magnitude of the parameters in
energy unit must be {\bf such that} $\sup \left\{ \Lambda _{QFT},\bar
{\mu}%
\right\} \ll \inf \left\{ \sigma \right\} $ with $\Lambda _{QFT}$
representing a general dimensional parameter (momenta or masses) in the 
QFT
in under consideration.

The preceding order automatically activates a limit operation with 
respect
to $\left\{ \sigma \right\} $ on the coarse grained amplitudes for
describing 'low' energy processes which will be denoted as $L_{\left\{
\sigma \right\} }\cdot \left( \equiv \lim_{\left\{ \sigma \right\}
\rightarrow \infty }\cdot \right) $. Then the coarse grained vacuum
functional in presence of the external sources for specifying low energy
processes reads
\FL
\begin{eqnarray}
&&Z\left( J\left( x\right) |\{\bar{c}\}\right) \equiv L_{\left\{
\sigma \right\} }Z\left( J\left( x\right) |\{\sigma
;\bar{\mu}\}\right),\nonumber \\ && Z\left( J\left( x\right)
|\{\sigma ;\bar{\mu}\}\right)\nonumber\\ &\equiv&
\int D\Phi \left( x{\bf |}\left\{ \sigma ;\bar{%
\mu}\right\} \right) \exp \left[ \frac i\hbar S\left( \Phi \left( x%
{\bf |}\left\{ \sigma ;\bar{\mu}\right\} \right) ;\left\{ \sigma ;\bar
{\mu}%
\right\} \Vert J\right) \right] ,\nonumber \\
\end{eqnarray}
where the $\left\{ \sigma ;\bar{\mu}\right\} $ dependence of a function
(al)
indicates that they are coarse grained objects well defined in QTOE. The
appearance of the constants $\{\bar{c}\}$ (including $\bar{\mu}$) in 
the RHS
of Eq.(1) implies that the order of functional integration and $L_
{\left\{
\sigma \right\} }$ can not be trivially exchanged, otherwise we would 
get
the ill defined QFT's or divergences, i.e.,
\begin{eqnarray}
&&L_{\left\{ \sigma \right\} }\int D\Phi \left( x{\bf |}\left\{
\sigma ;\bar{\mu}\right\} \right) \exp
\left[ \frac i\hbar S\left( \Phi \left( x{\bf |}\left\{ \sigma ;\bar{%
\mu}\right\} \right) ;\left\{ \sigma ;\bar{\mu}\right\} \Vert
J\right) \right]\nonumber \\ && \neq \int D\Phi \left( x\right)
\exp \left[ \frac i\hbar
S\left( \Phi \left( x\right) \Vert J\right) \right] ,
\end{eqnarray}
with $ S\left( \Phi \left( x\right) \Vert J\right) \equiv
L_{\left\{ \sigma
\right\} }S\left( \Phi \left( x{\bf |}\left\{ \sigma ;\bar{\mu}%
\right\} \right) ;\left\{ \sigma ;\bar{\mu}\right\} \Vert J\right) $ 
and $%
\Phi \left( x\right) \equiv L_{\left\{ \sigma \right\} }\Phi \left( x
{\bf |}%
\left\{ \sigma ;\bar{\mu}\right\} \right) $. In terms of Feynman diagram
algorithm, this is (for a one loop divergent diagram in QFT),
\begin{eqnarray}
&&L_{\left\{ \sigma \right\} }\Gamma ^{\left( 1-loop\right)
}\left( \left( p\right) ,\left( m\right) ;\{\sigma
;\bar{\mu}\}\right) \nonumber \\ &&\equiv L_{\left\{ \sigma
\right\} }\int d^DQ\bar{f}_\Gamma \left( Q,\left( p\right) ,\left(
m\right) ;\{\sigma ;\bar{\mu}\}\right) \nonumber \\ &&\neq \int
d^DQf_\Gamma \left( Q,\left( p\right) ,\left( m\right) \right),
\end{eqnarray}
with $f_\Gamma \left( Q,\left( p\right) ,\left( m\right) \right)
$denoting the integrand of the diagram defined in terms of usual
free propagators and vertices. The loop momentum, external momenta
and masses are denoted respectively by $Q,\left( p\right) $and
$\left( m\right) $.

In principle we could not evaluate the generating functional or the 
Feynman
amplitudes without knowing the exact dependence upon $\{\sigma ;\bar
{\mu}\}$%
. However, we can determine each one loop amplitude (ill defined in 
QFT) $%
L_{\left\{ \sigma \right\} }\Gamma ^{\left( 1-loop\right) }\left( \left(
p\right) ,\left( m\right) ;\{\sigma ;\bar{\mu}\}\right) $ up to an
appropriate polynomial of momenta and masses with finite but 
undetermined
coefficients {\bf as long as} we accept that the QTOE version of the 
loop
diagram exists\footnote{%
The worldlines of point particles in conventional QFT's should be 
replaced
by world-'volumes' (or world-'manifolds') in QTOE due to their 
complicated
underlying contents. An analogue can be found in string theories where 
the
particles' paths are worldsheets. Here we do not address whether the 
string
theories or M theory is the QTOE we refer to but focus on the benefits 
from
its existence.}.

{\bf THEOREM}. \hspace{0.01in}A one loop amplitude $\Gamma $ defined in 
QTOE
that corresponds to an ill defined one in conventional QFTs satisfies 
the
following kind of natural differential equation,%
\FL
\begin{eqnarray}
&&\left( {\partial }_p\right) ^{\omega _\Gamma +1}L_{\left\{ \sigma 
\right\} }%
\bar{\Gamma}\left( \left( p\right) ,\left( m\right) |\left\{ 
\sigma ;\bar{\mu%
}\right\} \right) \nonumber \\ &&=\int d^DQ\left( {\partial
}_p\right) ^{\omega _\Gamma +1}f_\Gamma (Q,\left( p\right) ,\left(
m\right) )\nonumber
\\
&& \equiv \Gamma ^{\left( \omega _\Gamma \right) }\left( \left(
p\right) ,\left( m\right) \right)
\end{eqnarray}
with $\omega _\Gamma $ being the superficial divergence degree or 
scaling
dimension of such a diagram and $f_\Gamma (Q,\left( p\right) ,\left(
m\right) )$ being the integrand of this diagram defined in conventional 
QFT.

${\sl Proof}$: Since QTOE is completely well defined,
then\cite{YYY} $$ \begin{array}{l}
 \left( {\partial }_p\right)^{\omega _\Gamma +1}
 L_{\left\{ \sigma \right\}}\bar{\Gamma}\left(
 \left( p\right) , \left( m\right) |\left\{\sigma ;\bar{\mu }\right\}
 \right) \\
 =L_{\left\{ \sigma \right\}}\int d^DQ
 \left( {\partial }_p\right) ^{\omega _\Gamma
+1}\bar{f}_{\bar{\Gamma}}(Q,\left( p\right) , \left(
m\right)|\left\{ \sigma ;\bar{\mu}\right\} )\\ =\int
d^DQ\left({\partial}_p\right)
^{\omega _\Gamma +1}L_{\left\{ \sigma \right\} }\bar{f}_{\bar{\Gamma}%
}(Q,\left( p\right) ,\left( p\right) |\left\{ \sigma
;\bar{\mu}\right\} )\\ =\int d^DQ\left( {\partial }_p\right)
^{\omega _\Gamma +1}f_\Gamma (Q,\left( p\right) ,\left( m\right)
)\\ \equiv \Gamma _d^{\left( \omega _\Gamma \right) }\left( \left(
p\right) ,\left( m\right) \right).\ \ \ \ Q.E.D.
\end{array}$$

Similar differential equations also hold with ${\partial }_p$ replaced 
by $%
\partial _m$. Solving such equations we get%
\FL
\begin{eqnarray}
&&L_{\left\{ \sigma \right\} }\bar{\Gamma}\left( \left( p\right)
,\left( m\right) |\left\{ \sigma ;\bar{\mu}\right\} \right)
\nonumber \\ &&\doteq \left( \int_p\right) ^{\omega _\Gamma
+1}\int d^DQ\left( {\partial }_p\right) ^{\omega _\Gamma
+1}f_\Gamma (Q,\left( p\right) ,\left( m\right) )
\end{eqnarray}
with the symbol '$\doteq $' indicating that the two sides are
equal up to certain integration constants in a polynomial of
momenta and masses of power $\omega _\Gamma $. To determine the
integration constants (which is definitely defined as
$\{\bar{c}\}$ in QTOE) we need 'boundary conditions' like
symmetries, sum rules and finally experimental data, which
parallels the procedure of choosing renormalization conditions.
Eq. (6) or (7) is just our general recipe for evaluating the
Feynman amplitudes that dispenses the notorious divergences and
the associated subtraction. This recipe works in the same way for
multiloop diagrams, for details please refer to Ref.\cite {YYY}.
The guideline is to insert a pair of $\left( \int_p\right)
^{\omega _\Gamma +1}$ and $\left( {\partial }_p\right) ^{\omega
_\Gamma +1}$ to the two sides of each divergent loop integration
as $L_{\left\{ \sigma \right\} } $ is moved across the loop
integration until the $L_{\left\{ \sigma \right\} }$ is finally
removed from all loops in the diagram. For convergent loops
$L_{\left\{ \sigma \right\} }$ can safely cross the loop
integrations. However, by defining that $\left( \partial \right)
^n\equiv {\left( \int \right) }^{-n},\ {\left( \int \right)
}^n\equiv \left( \partial \right)
^{-n},\ $for$\ n<0,{\left( \int \right) }^n=\left( \partial \right) 
^n=1,\ $%
for$\ n=0$ and noting that $\left( \int \right) ^n\times \left(
\partial \right) ^{n|}=\left( \partial \right) ^{-n}\times \left(
\int \right) ^{-n}=1 $ for $n<0$ we can also put a convergent loop
into the form of Eq.(7) with now $\omega _\Gamma $ denoting the
negative scale dimension of the convergent loop diagram.

We emphasize again that the above expressions are correct {\bf 
provided} the
magnitude order $\sup \left\{ |p|,m,\bar{\mu}\right\} \ll \inf \left\{
\sigma \right\} $ is satisfied, no matter how large the mass or 
momentum is.
It is also evident that our strategy is obviously applicable to any
interactions.

It is clear that in our strategy, no subtraction is necessary, no 
infinite
counterterms and bare parameters is present except finite 'bare'
parameters---the tree parameters in Lagrangian. Among the integration
constants (which will be denoted as $\left\{ C\right\} $ in contrast to 
$%
\left\{ \bar{c}\right\} $), there must be a dimensional scale to 
balance the
dimensions in the logarithmic function of momenta (which will be 
denoted as $%
\mu _{int}$ that corresponds to $\bar{\mu}$). The integration constants 
$%
\left\{ C\right\} $ span a space in which the QTOE prediction $\left\{ 
\bar{c%
}\right\} $ just lies on one point of this space. Obviously, the QTOE
definition of the Lagrangian constants and the 'loop' constants $\left\{
\bar{c}\right\} $ (including $\bar{\mu}$) should be scheme and scale 
(SAS)
invariant\cite{scheme,Maxw}. This observation hints us that if we start 
with
finite tree level Lagrangian constants then we should be able to pin 
down or
be close to the QTOE-determined constants $\left\{ \bar{c}\right\} $, 
as the
finite tree parameters are naturally 'bare' and hence SAS invariant. 
This
may accentuate and accelerate the extraction of physical parameters out 
of
renormalization scheme and scale dependent parametrization\cite{Maxw},
because if we could fix the Lagrangian parameters physically somehow
(through sum rules\cite{Narison}, for example), then we can 
approximately
extract $\left\{ \bar{c}\right\} $ from experiments. In this situation 
we
say that different choice of $\left\{ C\right\} $ would correspond to
different physics. (Note that the words 'tree or bare parameters' does 
not
mean no interaction, on the contrary, there are interactions--the 
classical
or tree interactions. In the QTOE picture, the tree parameters in fact
characterize the collective properties of the modes emerged out of the
underlying motions, which should be the same when the emergent modes 
undergo
annihilation and creation---the fluctuation of the these emergent 
quanta.)

From the preceding discussions on the constants $\{\bar{c}\}$, we can
parametrize them in such a way that $\{\bar{c}\}=\{\bar{\mu},\left[ \bar
{c}%
^0\right] \},$dim$\left\{ \bar{c}^0\right\} =0,\partial _{\bar{\mu}}\bar
{c}%
^0=\partial _g\bar{c}^0=0,\forall \bar{c}^0,\forall g:$dim$\left\{ 
g\right\}
\neq 0$. Then rescale every dimensional parameters in a general vertex
function $\Gamma ^{\left( n\right) }\left( \left( p\right) ,\left( 
g\right)
;\{\bar{c}\}\right) $ (we denote masses and couplings collectively as $%
\left( g\right) $) that is well defined in QTOE, we have 
\FL
\begin{equation}
\left\{ s\partial _s+\Sigma d_gg\partial _g+\bar{\mu}\partial _{\bar
{\mu}%
}-d_{\Gamma ^{(n)}}\right\} \Gamma ^{(n)}(\left( sp\right) ,\left( 
g\right)
;\{\bar{c}\})=0.
\end{equation}
with $d_{\cdots }$ denoting the mass dimensions of the associated
constants. Since all the constants $\{\bar{c}\}$ only appear in
the local parts of 1PI vertices, then $\bar{\mu}\partial
_{\bar{\mu}}$ induces the insertion of the operators
($\Sigma_{\{O\}}\delta _O\hat{I}_O$) corresponding to the
vertices,
that is,
\FL
\begin{eqnarray}
&&\bar{\mu}\partial _{\bar{\mu}}\Gamma ^{(n)}\left( \left(
p\right) ,\left( g\right) ;\{\bar{\mu},\left[ \bar{c}^0\right]
\}\right) \nonumber \\ &&=\Sigma_{\{O\}}\delta
_O\hat{I}_O\Gamma ^{(n)}\left( \left( p\right) ,\left( g\right) ;\{\bar
{\mu}%
,\left[ \bar{c}^0\right] \}\right) .
\end{eqnarray}
This is just the general form of renormalization group equation
(RGE) in our approach. Close investigation of the solutions of
Eq.(6) in terms of masses will show that the anomalous dimension
$\delta _O$ of vertex operator $O$
must be functions of dimensionless tree coupling $\left[ g^0\right] $ 
and $%
\left[ \bar{c}^0\right] $, i.e., $\delta _O=\delta _O\left( \left[
g^0\right] ,\left[ \bar{c}^0\right] \right) $\cite{YY}. The insertion 
of all
the Lagrangian operators with couplings $\left( g\right) $ can be 
realized
by $g\partial _g$ (for mass, it is ${m^k}\partial _{m^k},k=1(\text
{fermion}%
),2(\text{boson})$), i.e., $\Sigma_{\{O\}}\delta
_O\hat{I}_O=\Sigma \delta _gg\partial _g+\Sigma \delta _\phi
\hat{I}_{\partial \phi \partial \phi
}+\Sigma_{\{\bar{O}\}}\delta _{\bar{O}}\hat{I}_{\bar{O}},$ with $\phi $ 
and $%
\bar{O}$ denoting respectively the 'elementary' fields in Lagrangian 
and the
operators not defined in Lagrangian. Apparently $\Sigma_{\{\bar{O}\}}
\delta _{%
\bar{O}}\hat{I}_{\bar{O}}$ is absent in renormalizable theories,
while for unrenormalizable models, there will be infinitely many
$\bar{O}$ operators. The insertion of the kinetic operator $\delta
_\phi \hat{I}_{\partial \phi \partial \phi }$ will induce a
rescaling of the field operator $\phi $ by amount $\frac{\delta
_\phi }2$. Thus in renormalizable theories, we
obtain that 
\FL
\begin{equation}
\left\{ \bar{\mu}\partial _{\bar{\mu}}-\Sigma \bar{\delta}_gg\partial 
_g-%
\bar{\delta}_{\Gamma ^{\left( n\right) }}\right\} \Gamma ^{(n)}\left( 
\left(
p\right) ,\left( g\right) ;\{\bar{\mu},\left[ \bar{c}^0\right] \}
\right) =0
\end{equation}
with $\bar{\delta}_g\equiv \delta _g-\Sigma _{\left[ \phi \right] _g}
\frac{%
\delta _\phi }2$. Since $\left( g\right) $ and $\{\bar{\mu},\left[ \bar
{c}%
^0\right] \}$ should be uniquely determined by QTOE, the variation
in Eq.(10) should be understood as the change due to the global
rescaling of everything. Thus by introducing a natural set of
scale co-moving (or 'running') parameters basing on Coleman's
bacteria analogue\cite{Cole}, we
finally arrive at the standard form of RGE%
\FL
\begin{equation}
\left\{ \mu \partial _\mu -\Sigma \bar{\delta}_{\bar{g}}\bar{g}\partial 
_{%
\bar{g}}-\bar{\delta}_{\Gamma ^{\left( n\right) }}\right\} \Gamma
^{(n)}\left( \left( p\right) ,\left( \bar{g}\right) ;\{\mu ,\left[ \bar
{c}%
^0\right] \}\right) =0,\ \ \
\end{equation}
with $\mu \partial _\mu \bar{g}\left( \mu ;\left( g\right) \right) =\bar
{g}%
\left( \mu ;\left( g\right) \right) \bar{\delta}_{\bar{g}}\left( \left[ 
\bar{%
g}^0\left( \mu ;\left[ g^0\right] \right) \right] ,\left[ \bar{c}^0
\right]
\right) ,$ $\ \bar{g}\left( \mu ;\left( g\right) \right) |_{\mu =\bar
{\mu}%
}=g,\ \ \mu \equiv t\bar{\mu},\ \hspace{0.01in}t:\max \left[ \mu
\right] \ll \inf \left\{ \sigma \right\} $. Now we see that the
'running' of the parameters is closely related to the coarse
graining procedure and the clustering phenomenon, the mystery
atmosphere around the dimensional transmutation phenomenon is
therefore removed. Only $\mu=t\bar{\mu}$ runs, while $\bar{\mu}$
specifies the physical reference scale for coarse graining. All
the subtleties mentioned in the beginning disappeared here as
there is no divergence and hence no infinite bare parameters in
our derivation.

Inserting Eq.(11) back into Eq.(8) we will get the full scaling
equation due to Callan-Symanzik\cite{CS}
\FL
\begin{eqnarray}
&&\left\{ s\partial _s+\Sigma \bar{\delta}_{\bar{g}}\bar{g}\partial _
{\bar{g}}+%
\bar{\delta}_{\Gamma ^{\left( n\right) }} -d_{\Gamma
^{(n)}}\right\} \Gamma ^{(n)}\left( \left( sp\right) ,\left(
\bar{g}\right) ;\{\bar{\mu},\left[ \bar{c}^0\right]
\}\right)\nonumber \\ &&
=-i\Gamma _\Theta ^{(n)}\left( 0,\left( sp\right) ,\left( \bar{g}
\right) ;\{%
\bar{\mu},\left[ \bar{c}^0\right] \}\right) ,
\end{eqnarray}
where
\FL
\begin{eqnarray}
&&s\partial _s\bar{g}\left( s\bar{\mu};\left( g\right) \right)
 =\bar{g}\left( s\bar{\mu};\left( g\right) \right) \bar{\delta}_{\bar
{g}}\left(
\left[ \bar{g}^0\left( s\bar{\mu};\left[ g^0\right] \right)
\right] ,\left[ \bar{c}^0\right] \right),\\ &&
\bar{g}\left(s\bar{\mu};\left( g\right) \right) |_{s=1}=g,\\ & &
i\Gamma _\Theta ^{(n)}\left( 0,\left( sp\right) ,\left(
\bar{g}\right) ;\{\bar{\mu},\left[ \bar{c}^0\right]
\}\right)\nonumber \\ &&\equiv \Sigma d_{\bar{g}}\bar{g}\partial
_{\bar{g}}\Gamma ^{(n)}\left( \left( sp\right) ,\left(
\bar{g}\right) ;\{\bar{\mu},\left[ \bar{c}^0\right] \}\right) ,
\end{eqnarray}
with $\Theta $ being the trace of the energy
tensor of the theory. Of course in reality we are forced to replace 
$\{\bar{c%
}\}$ with $\{C\}=\{\mu _{int},\left[ C^0\right] \}$\footnote{Here
$\mu _{int},\left[ C^0\right]$ parallel $\bar{\mu},\left[ \bar{c}%
^0\right]$ with $\mu _{int}$ standing for the dimensional constant
scale that will necessarily appear in the indefinite momentum
integration and $\left[ C^0\right]$ for those dimensionless
constants}, but in principle we can start with tree parameters and
determine $\{C\}$ by confronting our calculations with
experimental data as mentioned above. Moreover, if we start with
the tree parameters (SAS invariant), then the so-called scheme
dependence problem can be converted into the problem of assigning
the tree
parameters physical values somehow and extracting the QTOE information 
$\{%
\bar{c}\}$ (in an approximate way, of course) from experiments.

Conventionally, we start with infinite bare parameters (due to bad
regularization schemes in use) to 'construct' renormalized ones that 
run.
Then we try to reexpress the \hspace{0.01in}running parameters that are
scheme dependent in terms of a set of parameters that are both scale and
scheme independent\cite{scheme,Maxw}. One might ask that, why do we 
work so
awkwardly? Why do not we just start from a set of scale and scheme
independent parameters? No good answer to such a simple query can be 
found
within the conventional theory, as removing divergences is the first
annoying task. While our plausible and simple strategy automatically
dispensed the long standing subtleties in conventional renormalization
programs with all their merits kept and improved. We also clarified the
origin of the problem and the subsequent difficulties in the 
conventional
approaches.

One might oppose that the QTOE is never seen. Our answer might be that 
the
present formulation or field equation has never been verified both
theoretically and experimentally in the high energy limit, why should 
one be
so confident in extrapolating the field equations to the ranges where 
they
have never been justified? As a matter of fact, it is no harm to start 
with
a postulated underlying regularity, if there is no need of this 
property,
then one can find that, the assumption can be freely removed at any 
time.
While in the present QFT's, we do need such underlying regularity.

Before ending our presentation, we note that RGE is in fact a
decoupling theorem of the underlying high energy modes since if we
do not take the low energy limit then we can not expand the
contribution from the variation of the coarse graining threshold
in terms of the low energy operators in Eq.(9)
and Eq.(8) must be written as
\FL
\begin{eqnarray}
&&\left\{ s\partial _s+\Sigma d_gg\partial _g+\Sigma d_\sigma \sigma 
\partial _\sigma -d_{\Gamma ^{(n)}}\right\} \bar{%
\Gamma}^{(n)}(\left( sp\right) ,\left( g\right) ;\{\sigma
;\bar{\mu}\}) \nonumber \\ &&=0.
\end{eqnarray}
Thus the physical meaning of RGE is deepened in the new strategy.
Note that the $ \bar{\mu}\partial _{\bar{\mu}}$ is absent here
since the vertex function is totally well-defined in QTOE or with
$\{\sigma\}$ present, i.e., the coarse graining reference scale
just serves as a separating scale for performing quantum loop
calculations: separating those modes typically characterized by
$\{\sigma\}$ and those characterized by $[g]$ (Lagrangian
parameters for effective theories). Only when the underlying
constants are vanishingly small do we need $\bar{\mu}$ to stand
for the underlying structures' influences. We emphasize again that
our approach naturally requires a threshold scale for each QFT or
effective sectors to coarse grain the short distance processes,
therefore the mystery in the dimensional transmutation phenomenon
is removed. We also remind that the present formulation are only
valid provided the underlying modes' typical time scale is
vanishingly small comparing to the QFT processes' time scale. In
this sense the parameters in QFT's are some kind of collective
'coordinates' of the coarse grained clusters. So our approach is
in fact pointing towards a unified framework in which
quantization, coarse graining, renormalization and unification of
interactions are coherently organized.

Finally, we mention again that our approach accentuates the importance 
of
fixing the ambiguities or the choice of renormalization conditions, 
which
are especially important for the complicated electroweak theory with
spontaneous symmetry breaking\cite{EW}. Further applications of our 
simple
and natural strategy to various problems involved with divergence and/or
singularity will be interesting.

In summary, we demonstrated a natural strategy for renormalization as a
refinement of the conventional approaches, which automatically 
dissolves the
long standing subtleties. The key point is the existence of \hspace
{0.01in}a
complete quantum theory of everything which contains full information 
of the
high energy limit with the QFT's correspond to the 'low' energy sectors 
of
the QTOE. The necessary coarse graining of the short distance details
naturally requires a threshold scale and the physical origin 
of 'running' is
closely related to this mechanism. From this QTOE scenario, there 
descends a
natural series of differential equations for evaluating the Feynman 
diagrams
as an alternative technical approach.


\begin{references}
\bibitem{Wilson}  K. G. Wilson and J. Kogut, Phys. Rep. {\bf C12}, 75
(1974).

\bibitem{RMP}  See, e.g., L. R. Surguladze and M. A. Samuel, Rev. Mod. 
Phys.
{\bf 68}, 259 (1996).

\bibitem{Cluster}  K. Wilson, Phys. Rev. Lett. {\bf 27}, 690 (1971); J.
Kogut and L. Susskind, Phys. Rev. {\bf D9}, 697, 3391 (1974).

\bibitem{Coarse}  R. B. Griffiths, J. Stat. Phys. {\bf 36}, 219 (1984),
Phys. Rev. Lett. {\bf 70}, 2201 (1993); M. Gell-Mann and J. B. Hartle, 
Phys.
Rev. {\bf D47}, 3345 (1993); R. Omnes, Rev. Mod. Phys. {\bf 64}, 339 
(1992).

\bibitem{YYY}  Ji-Feng Yang, Report No. hep-th/9708104; invited talk,
pp202-206 in {\it Proceedings of the XIth International
Conference:''Problems of Quantum Field Theory'98(Dubna)''}. Eds: B. M.
Barbashov, G. V. Efimov and A. V. Efremov. Dubna, Russia,1999; Report 
No.
hep-th/9904055.

\bibitem{scheme}  P. M. Stevenson, Phys. Rev. {\bf D23}, 2916 (1981); G.
Grunberg, Phys. Rev. {\bf D29}, 2315 (1984); S. J. Brodsky, G. P. 
Lepage and
P. B. Mackenzie, Phys. Rev. {\bf D28}, 228 (1983).

\bibitem{Maxw}  D. T. Barclay, C. J. Maxwell and M. T.\hspace{0.01in}
Reader, Phys. Rev. {\bf D49}, 3480 (1994).

\bibitem{Narison}  S. Narison, {\it QCD Spectral Sum Rules}, Lecture 
Notes
in Physics 26 (World Scientific, 1989).

\bibitem{YY}  Ji-Feng Yang, Report No. hep-th/0005195.

\bibitem{Cole}  S. Coleman, {\it Aspects of Symmetry}, (Cambridge 
University
Press, Cambridge, England, 1985), Chapter 3.

\bibitem{CS}  C. G. Callan, Jr., Phys. Rev. {\bf D2}, 1541 (1970); K.
Symanzik, Comm. Math. Phys. {\bf 18}, 227 (1970).

\bibitem{EW}  See, e.g., B. A \hspace{0.01in}Kniehl and A. Sirlin, 
\hspace{%
0.01in}Phys. Rev. Lett. {\bf 81}, 1373 (1998); M. Passera and A. Sirlin,
Phys. Rev. {\bf D58, }113010 (1998) and references therein.
\end{references}
\end{document}